\documentclass[runningheads]{llncs}
\usepackage{graphicx}
\usepackage{listings}

\begin{document}
\title{CARAVAN: a framework for comprehensive simulations on massive parallel machines}
\titlerunning{CARAVAN: a framework for comprehensive simulation}
%
\author{
Yohsuke Murase\inst{1}\orcidID{0000-0002-5872-420X} \and 
Hiroyasu Matsushima\inst{2}\orcidID{0000-0001-7301-1956} \and
Itsuki Noda\inst{3}\orcidID{0000-0003-1987-5336} \and
Tomio Kamada\inst{1,4}\orcidID{0000-0002-1646-1683} 
}
\authorrunning{Y.~Murase et al.}
%
\institute{RIKEN Center for Computational Science, Kobe, Hyogo 650-0047, Japan
\email{yohsuke.murase@gmail.com}\\
\and
School of Engineering, The University of Tokyo, Bunkyo-ku, Tokyo 113-8656, Japan\\
\email{matsushima@sys.t.u-tokyo.ac.jp}
\and
Artificial Intelligence Research Center, AIST, Tsukuba, Ibaraki 305-8560, Japan\\
\email{i.noda@aist.go.jp}
\and
Grad. School of Sys. Informatics, Kobe University, Kobe, Hyogo 657-8501, Japan\\
\email{kamada@fine.cs.kobe-u.ac.jp} 
}
\maketitle              
\begin{abstract}
We present a software framework called CARAVAN, which was developed for comprehensive simulations on massive parallel computers.
The framework runs user-developed simulators with various input parameters in parallel without requiring the knowledge of parallel programming.
The framework is useful for exploring high-dimensional parameter spaces, for which sampling points must be dynamically determined based on the previous results.
Possible use cases include optimization, data assimilation, and Markov-chain Monte Carlo sampling in parameter spaces.
As a demonstration, we applied CARAVAN to an evacuation planning problem in an urban area.
We formulated the problem as a multi-objective optimization problem, and searched for solutions using multi-agent simulations and a multi-objective evolutionary algorithm, which were developed as modules of the framework.

\end{abstract}
\section{Motivation and Significance}

The advancement of information and communication technologies in recent decades revolutionized the study of social behavior as we gained access to the huge number of the digital records, so called ``big-data'', of our daily activities.
Various mathematical methods and algorithms have successfully been applied to analyze these empirical data to characterize the societal activities.
After empirical verification of the data, the next steps are model development and its simulations to deepen our understanding of the underlying mechanisms.
Multi-agent social simulation (MASS) serves as a powerful tool because social systems often demonstrate non-trivial collective phenomena that emerge from the actions of individuals, including the occurrence of traffic jams, bursty spreading of rumors on social networks, and a sudden crash in economic markets.
Through the development of the models which based on the descriptions at an individual level, we are able to study the causal relationships between microscopic activities and their emergent macroscopic consequences.
Moreover, well-developed MASS is expected to contribute to the better design of our social systems and services through the simulation of various possible future scenarios.

However, as discussed in~\cite{noda_jcss2018}, the application of MASS is not as straightforward as that of simulations for physical systems.
One of the most critical difficulties is the fact that models for MASS are not as well established as those for physical systems.
Models for MASS inevitably involve a non-negligible amount of uncertainty because individual behavior is the outcome of highly complicated intellectual, psychological, and behavioral processes that are different for each person.
Furthermore, multiple social phenomena, such as the economy and traffic, may mutually interact, which makes it even more difficult to identify the factors to incorporate into a model.
Even big data cannot be a solution to these problems because the data are often incomplete and biased~\cite{torok_pre2016} because of technical and privacy issues.

One of the methodologies to overcome these difficulties is the use of an exhaustive simulation~\cite{noda_jcss2018}.
By its nature, it is impossible to precisely predict an actual social system using a single run of MASS.
Instead, it is more productive to investigate the global phase diagram of the system by running simulations with various assumptions and parameters to compensate for uncertainty.
Such exhaustive simulations require both a huge amount of computational resources and effective algorithms to explore broad parameter spaces; hence, the effective application of high-performance computers (HPCs) are necessary.

In~\cite{noda_jcss2018}, Noda et al. discussed the expected computational scales for several domains of MASS and summarized them as a road map.
According to the road map, although it is hypothetical, the number of required runs for a research issue in the coming decades will be order of $10^2 \sim 10^6$.
Although it is a so-called embarrassingly parallel problem, running such a large number of simulation jobs is not a simple issue from a technical point of view.
Furthermore, intelligent algorithms for sampling parameter spaces are required as a naive random sampling would evidently be useless in a high-dimensional space.
Hence, software frameworks are needed in order to correctly manage an enormous number of jobs on massive parallel computers, and to provide functions to define workflow to sample parameter spaces effectively.
One of the solutions to address this problem is software called ``OACIS'', which manages simulation jobs automatically and provides a simple interface for users~\cite{murase_jphys2017,murase_physproc2014}.
Although OACIS works fairly well for a wide range of problems, it can only manage up to $10^2 \sim 10^4$ jobs because of the design decision to maximize usability and versatility.
To manage even more jobs easily, we need another framework that is more specialized in terms of scalability.

In this article, we present a software framework called CARAVAN for parameter-space exploration on massive parallel super-computers.
It was developed as an open-source software and is available on github~\cite{caravan-github}.
By combining a simulator developed by a user with CARAVAN, we are able to run the simulator with various input parameters in parallel, making full use of HPCs.
As shown in the next section, it scales well up to tens of thousands of processes and can manage millions of tasks.
Using the framework, users become free from writing a code for parallelization using an Message Passing Interface (MPI) library because concurrent execution and scheduling of the simulation are managed by the framework.
Furthermore, it is applicable not only to trivial parameter parallelization but to more complex parameter searching, such as optimization or Markov chain Monte Carlo sampling, for which sampling points are dynamically determined based on the previous results.
In the next section, we illustrate the architecture of CARAVAN.
Details of the implementation and its performance evaluation on the K computer are shown in Section~\ref{sec:implementation}.
In Section~\ref{sec:example}, we present the application of CARAVAN to a MASS for evacuation guidance.
In the final section, we present a summary and future perspectives.

\section{Software Description}

\subsection{Overall Architecture}

\begin{figure}
  \begin{center}
  \includegraphics[width=.9\textwidth]{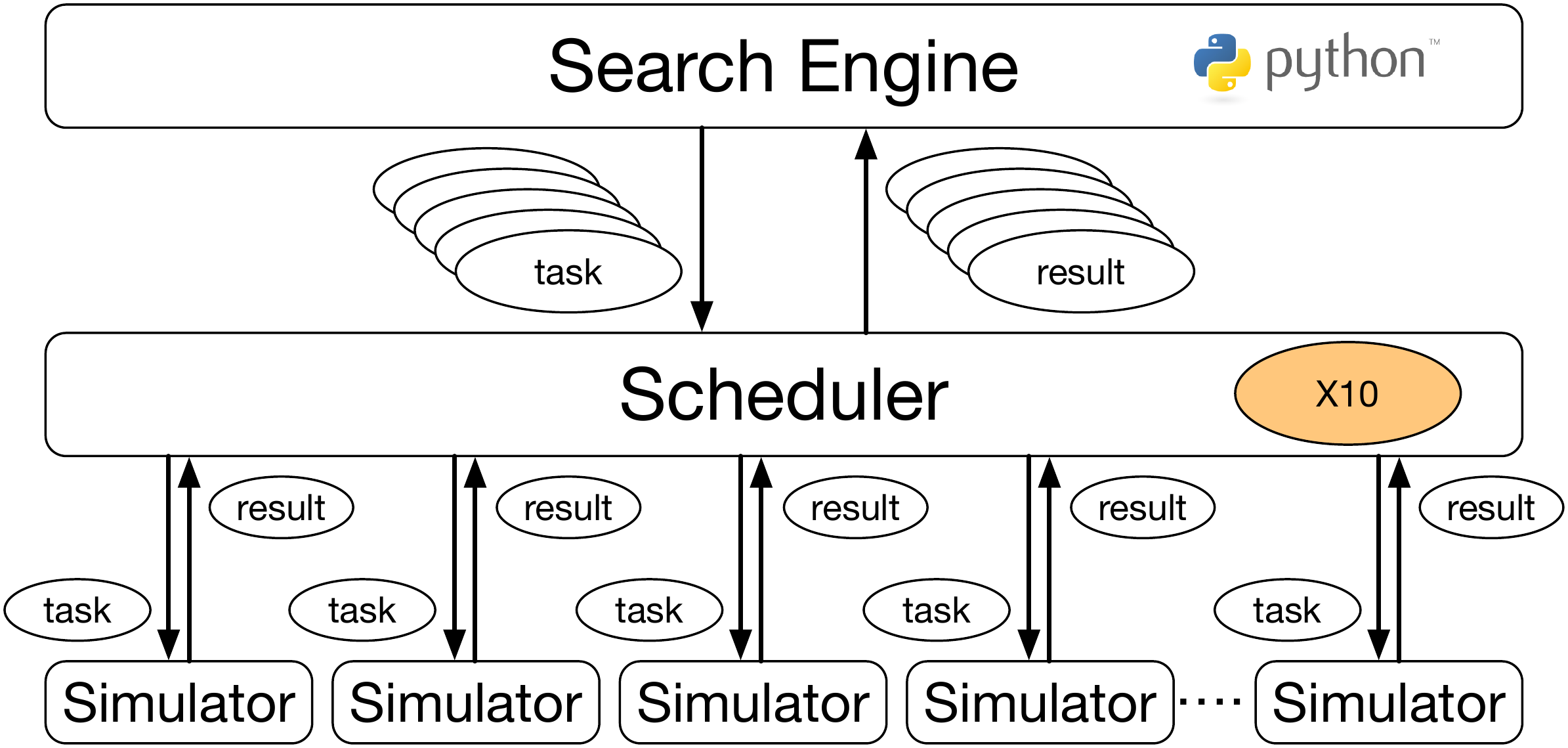}
  \caption{
  An overview of the architecture of CARAVAN.
  \label{fig:overview}
  }
  \end{center}
\end{figure}

Figure~\ref{fig:overview} illustrates the architecture of CARAVAN.
It consists of three modules: ``search engine,'' ``scheduler,'' and ``simulator.''

\begin{description}
\item[Simulator]
A simulator is an executable application that the user wants to execute in parallel.
It is executed as an external process by the scheduler that receives input parameters as command line arguments.
A single execution of a simulator is called a ``task'' in CARAVAN.
\item[Scheduler]
The scheduler is the module that is responsible for parallelization.
It receives commands to execute simulators from the search engine, distributes them to available nodes, and executes the simulators in parallel.
This module is implemented in the X10 language~\cite{x10-lang}, which is compiled into native code linked to an MPI library.
\item[Search Engine]
The search engine is a module that determines the policy on how parameter space is explored.
More specifically, it generates a series of commands to be executed, that is tasks, and sends them to the scheduler.
When a task is complete, the search engine receives its results from the scheduler.
Based on the received results, the search engine can generate other series of tasks repeatedly.
Because tasks are executed in parallel, communication between the search engine and the scheduler occurs asynchronously.
\end{description}

Among the three modules, users prepare a simulator and a search engine to conduct parameter-space exploration.
A simulator is implemented as an executable program to be integrated into the framework.
Because it is an external process, a user can implement a simulator in any language.

A search engine is the module to define the workflow of parameter-space sampling.
Because parameter space is usually a high-dimensional space, various types of importance sampling, such as evolutionary optimization or Markov-chain Monte Carlo sampling, must be conducted.
Hence, the parameter space to explore must be dynamically determined based on the existing simulation results, which is hard to realize with a Map-Reduce like framework.
To implement such sampling algorithms, the framework provides a set of Python functions, or application programming interfaces (APIs), including ones to define callbacks which are invoked when tasks are complete.

The scheduler module is not modified by users; therefore, users do not have to write any X10 code by themselves.
Once a simulator and a search engine are implemented, users can conduct parameter space exploration using tens of thousands of processors.


\subsection{Requirements for a Simulator}

A simulator is a stand-alone executable program that must satisfy the following requirements:

\begin{itemize}
  \item accept parameters for simulations as command line arguments;
  \item generate outputs in the current directory; and
  \item (optional) write results to the ``\_results.txt'' file.
\end{itemize}

First, a simulator must be prepared such that it accepts input parameters as command line arguments.
This is because the scheduler receives a series of command lines from the search engine and executes them as an external process.
Another requirement for a simulator is that it must generate its output files or directories in the current directory.
This is because the scheduler creates a temporary directory for each task and invokes the command after setting the temporary directory as the current directory.

If a user's simulator writes a file called ``\_results.txt,'' it is parsed by the scheduler and its contents are sent back to the search engine.
This is useful when a user's search engine determines the next parameters according to the simulation results.
For instance, if users would like to optimize a certain value of the simulation results, they should write a value that they want to minimize (or maximize) to the ``\_results.txt'' file.
The file may contain several floating point values as its result.

\subsection{Preparation of a Search Engine}

The search engine is responsible for generating the command to be executed by the scheduler.
An example of a minimal program for the search engine is as follows:

\begin{lstlisting}
import sys
from caravan.server import Server
from caravan.task import Task

with Server.start():
    for i in range(10):
        Task.create("echo hello caravan %d" % i)
\end{lstlisting}

This sample creates a list of tasks, each of which runs the \lstinline{echo} command.
These commands are distributed to the subprocesses of the scheduler and executed in parallel.

In many applications, such as optimization, new tasks must be generated based on the results of completed tasks.
Methods to define callback functions are provided for that purpose:

\begin{lstlisting}
with Server.start():
    for i in range(10):
        task = Task.create("sleep %d" % (i%3+1))
        task.add_callback(lambda t, ii=i: Task.create("sleep %d" % (ii%3+1)))
\end{lstlisting}

If users run this program, they will find that 10 tasks are created, and 10 more tasks are created after each of the initial tasks is completed.

Although callbacks work fine, the code tends to become too complicated because of deeply nested callbacks.
One of the best practices to avoid complexity is to use a ``async/await'' pattern, for example,

\begin{lstlisting}
def run_sequential_tasks(n):
    for t in range(5):
        task = Task.create("sleep %d" % ((t+n)%3+1))
        Server.await_task(task)
        # this method blocks until the task is finished.

with Server.start():
    for n in range(3):
        Server.async( lambda n=n: run_sequential_tasks(n) )
\end{lstlisting}

This program spawns three concurrent activities, each of which executes five tasks sequentially.
For each activity, a new task is created after the previous task is complete.
If users visualize the results of the following program, they will see three concurrent lines of sequential tasks of length five.

In addition to the ``await'' method, the ``await\_all\_tasks'' method is also provided to wait for a set of tasks to complete.
After awaiting tasks, users can obtain the results of the simulation runs by accessing the ``results'' attribute of the task.
Using these methods, users can achieve a program in which tasks are created depending on the results of completed tasks.

There are also other classes and methods, such as ``ParameterSet'' and ``Run,'' to simplify the implementation of Monte Carlo sampling.
We do not present the full list of the APIs here. For the full documentation, please refer to the repository of CARAVAN~\cite{caravan-github}.

\section{Implementation}\label{sec:implementation}

CARAVAN as a whole is executed as a single MPI job.
When the MPI process starts, the rank $0$ process (hereafter, the root process) invokes a Python process of the search engine as an external process.
The search engine process communicates with the root process using bidirectional pipelines, thereby sending the information of simulation tasks and receiving their results.
Once a series of tasks is sent to the root process, they are distributed to the other subprocesses via an MPI protocol, that is, these MPI processes work as the scheduler module.
The subprocesses that receive the tasks then call the simulator, and wait until its simulation is complete.
The results are parsed by the subprocesses of the scheduler, and then sent back to the search engine.

CARAVAN was designed for cases in which the duration of each task (a single run of user's simulator) typically ranges from several seconds to a few hours.
CARAVAN does not perform quite well for tasks that are complete in less than a few seconds.
One of the reasons for this limitation originates from the design decision that a simulator is executed as an external process.
For each task, CARAVAN creates a temporary directory, creates a process, and reads a file generated by the simulator, which represents some overheads.
If users would like to run fine-grained tasks, they should consider using Map-Reduce frameworks, such as \cite{kmapreduce2013}.
Instead, the CARAVAN scheduler is designed such that it achieves ideal load balancing, even when the durations vary by orders of magnitude.
Tolerance for a variation in time is essential for parameter space exploration because elapsed times typically depend significantly on the parameter values.
CARAVAN was designed to scale up well to tens of thousands of MPI processes for tasks of this scale.

\begin{figure}
  \begin{center}
  \includegraphics[width=.85\textwidth]{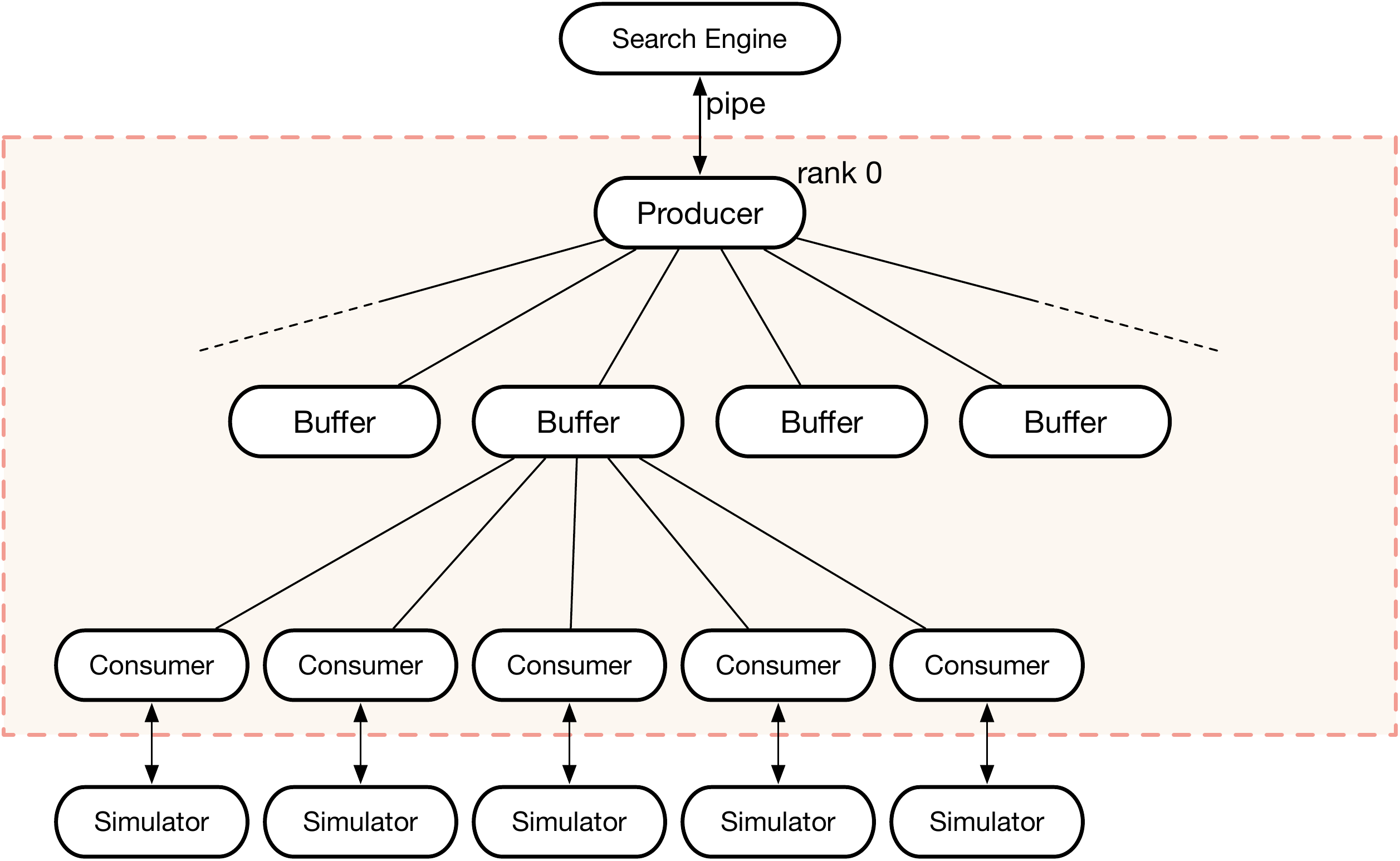}
  \caption{
  The internal design of the scheduler module.
  Each rounded rectangle corresponds to a process.
  The shaded area denotes the scheduler module, which is implemented as MPI processes.
  The producer, which is a rank $0$ MPI process, communicates with the search engine via bidirectional pipes.
  Tasks are distributed to buffers and then sent to their producers.
  Each consumer spawns a simulator process as its subprocess.
  }
  \label{fig:buffers_in_scheduler}
  \end{center}
\end{figure}

The scheduler module adopts the producer-consumer pattern, but with a ``buffered'' layer between the producer and its consumers, as shown in Fig.~\ref{fig:buffers_in_scheduler}.
The root process works as a producer.
The producer has hundreds of buffer processes, each of which has hundreds of consumer processes.
The buffered layer is introduced to prevent communication overload in a massive parallel environment.
Without the buffered layer, the producer process must communicate with thousands or more consumer processes, which causes technical problems and the entire process cannot be completed normally.
By introducing the buffered layer, the producer communicates only with hundreds of buffer processes.
The buffer processes have their own task queues to store the tasks, and repeatedly send them to their consumers gradually, significantly reducing the amount of communication of the producer process.
A similar mechanism is also adopted for the other direction of communication.
The buffer processes have a store to keep the results for a short time to prevent too frequent communication.
By default, CARAVAN allocates one buffer process to $384$ MPI processes, which is a good parameter for a wide range of practical use cases.

The current version of CARAVAN supports only serial or multi-thread parallel programs as simulators.
It cannot invoke an MPI-parallelized program as a simulator because CARAVAN launches the simulation command as an external process using a ``system'' command, not as an MPI process invoked by an ``MPI\_Comm\_Spawn'' function.
In a future release, we plan to support MPI-parallelized simulators.

We evaluated the performance of job scheduling for the following test cases:

\begin{description}
  \item[case 1 (TC1)]
  At the beginning of the job, we generate $N$ tasks.
  Each task takes $t$ seconds, where $t$ is drawn randomly from a uniform distribution $[20, 30]$.
  \item[case 2 (TC2)]
  At the beginning of the job, we generate $N$ tasks, whose duration $t$ is drawn from a power law distribution of exponent $-2$ between $t_{\rm min}=5$ and $t_{\rm max}=100$ seconds.
  \item[case 3 (TC3)]
  At the beginning of the job, we generate $N/4$ tasks.
  When each task is complete, another task is created until the total number of tasks reaches $N$.
  The duration of each task $t$ is drawn randomly from a power law distribution of exponent $-2$ between $t_{\rm min}=5$ and $t_{\rm max}=100$ seconds.
\end{description}

TC1 corresponds to the case in which the variation in task durations is not large.
This is the easiest among the three cases because its load balancing is trivial.
TC2 is more complicated because the distribution of the task durations has a heavy tail.
The majority of the jobs are complete in less than $10$ seconds; however, there are a certain number of tasks that run for significantly longer durations.
TC3 is even more complicated because all tasks are not generated initially. Tasks are appended after the jobs are complete.
We test this case because we often need to determine the parameter space to be explored depending on the results of previous tasks.
For these tests, we generated dummy tasks, each of which slept for a given period of time.

\begin{figure}
  \begin{center}
  \includegraphics[width=.85\textwidth]{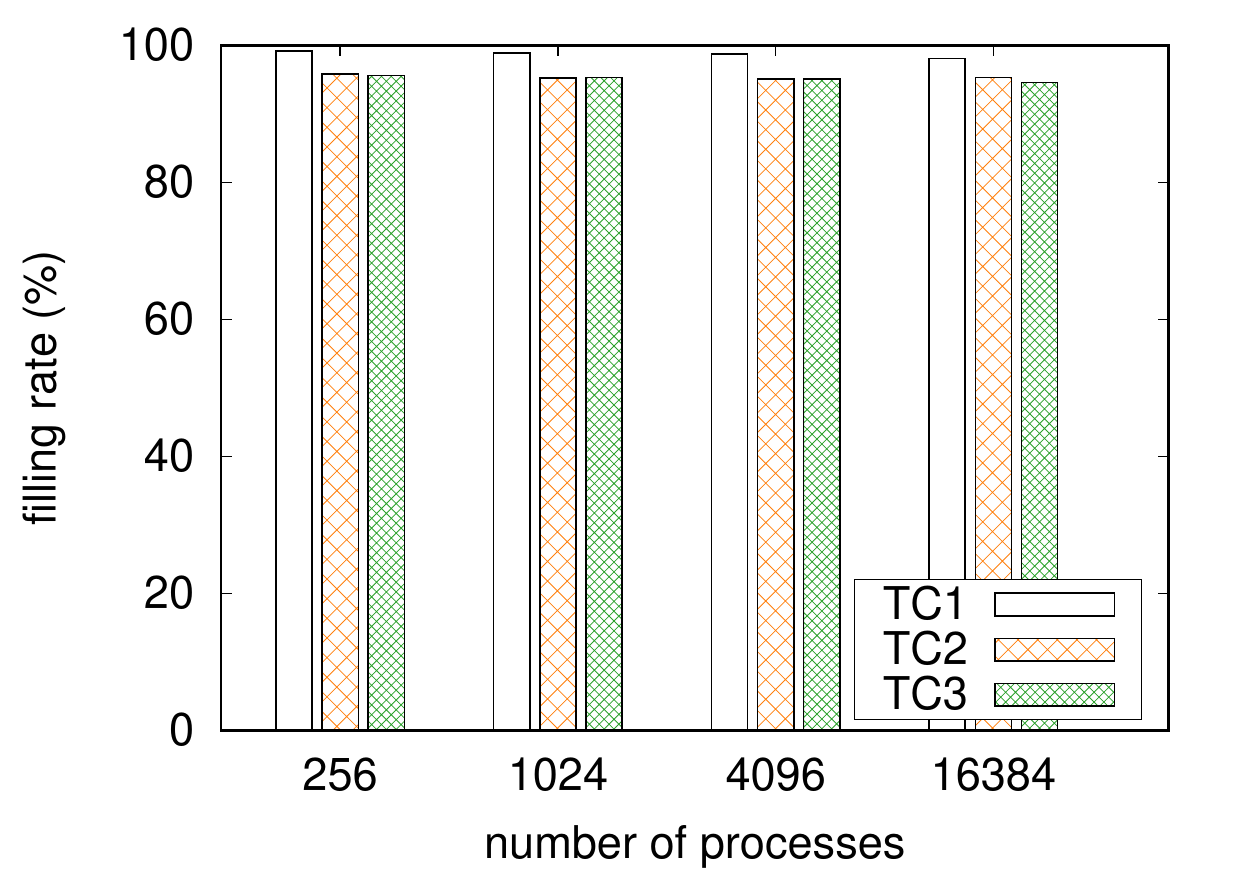}
  \caption{
  Performance of the CARAVAN for the three test cases on the K computer.
  The job filling rates for TC1, TC2, and TC3 are depicted for several numbers of MPI processes.
  }
  \label{fig:scaling}
  \end{center}
\end{figure}

We evaluated these test cases on the K computer using $N_{p} = 256$, $1024$, $4096$, and $16384$ MPI processes.
The number of nodes used in these tests was $N_{p}/8$ because a node of the K computer has eight cores and the tests were conducted as flat-MPI jobs.
We used $N=100N_{p}$; hence, each MPI process had $100$ tasks, on average.
We evaluated the performance using the job filling rate $r$, which we define as
\begin{equation}
  r = \frac{\sum_{i}^{N}\left(t_i^{\rm end}-t_i^{\rm begin}\right) }{T*N_{p}},
\end{equation}
where $t_i^{\rm end}$ and $t_i^{\rm begin}$ are the times at which the $i$th task begins and ends, respectively.
Total job duration $T$ is defined as the interval between the beginning of the first task and end of the last task, that is, $T = \max\{t_i^{\rm end}\} - \min\{t_i^{\rm begin}\}$.
The job filling rate is an indicator of the equal load balancing and the cost of inter-process communications.
If the communication cost is negligible and the load is perfectly balanced, the job filling rate should reach $100\%$.
The results of the performance evaluation on the K computer are shown in Fig.~\ref{fig:scaling}.
As shown in the figure, the job filling rates for the three test cases were reasonably close to the optimum, which demonstrates ideal scaling up to this scale.

\section{Application to Multi-Agent Simulation}\label{sec:example}

\subsection{Searching Trade-Off Relationships in Evacuation Planning}
Designing a response plan to disasters is not a simple optimization problem.
For example, when designing an evacuation plan for residents, we need to optimize its effectiveness (e.g. duration to complete the evacuation) while taking into account its feasibility and cost.
Even a highly effective plan cannot be adopted when it requires an infeasible cost to be implemented.
There often exist trade-offs between these factors; thus, planning a disaster response can be formulated as a multi-objective optimization problem.

In this section, as a case study, we investigate the trade-off relationships of evacuation plans for a flood caused by a tsunami in a district in Japan.
We use a multi-objective evolutionary algorithm (MOEA)~\cite{deb:moea} to locate the Pareto front in three-dimensional space of the effectiveness, cost, and feasibility, where these values for each plan are estimated using a MASS.
(Details of the objective functions are provided later.)
An MOEA is implemented on CARAVAN because it requires many simulation runs with various evacuation plans.

\subsection{Multi-Objective Optimization Algorithm}

Multi-objective optimization involves optimizing more than one objective function simultaneously, where a number of Pareto optimal solutions exist in general.
It is formulated as
\begin{equation}
  \min \left( f_{1}(\mathbf{x}), f_{2}(\mathbf{x}), \dots, f_{k}(\mathbf{x}) \right),\\
  \label{eq:objective}
\end{equation}
where $k$ is the number of objective functions and $f_{i}$ is the $i$th objective function of a set of variables, $\mathbf{x}$.
An MOEA is a variant of the evolutionary algorithm for multi-objective optimization problems, which repeats (1)~parent selection, (2)~crossover, (3)~mutation, and (4)~deletion to update the population.
In this cycle, the MOEA retains good solutions in the previous generation as archived solutions.
We adopt one of the most standard methods of an MOEA, the elitist non-dominated sorting genetic algorithm NSGA-II~\cite{deb:nsga}.

In the conventional NSGA-II, a population update is performed after the objective functions for all the individuals in the population have been calculated, that is, after multi-agent simulations that correspond to all individual cases are completed in our case.
Although we can evaluate objective functions in parallel using HPCs, a naive implementation of NSGA-II may cause serious performance degradation.
This is because the times required to run simulations for these individual cases may be widely different.
If we wait for the completion of the calculations for all individuals, a significant amount of CPU resource is wasted because of the serious load imbalance.

To overcome this problem, we introduce an asynchronous generation-update method to NSGA-II.
In our algorithm, we update a subset of the population when a certain fraction of the calculations are complete without waiting for all the simulation runs to be completed.
More specifically, we prepare $P_{\rm ini}$ individuals at the beginning and start calculations for them.
When the calculations for $P_{n}$ ($< P_{\rm ini}$) individuals are complete, they are added to the set of archived individuals.
Based on the results of the archived individuals, $P_{n}$ offspring are newly generated and calculations for them are started.
This replacement of $P_{n}$ individuals is defined as a single generation, and we repeat this process for a given number of generations.
When $P_{n}$ newly complete individuals are added to the archived individuals, we keep only the top $P_{\rm archive}$ individuals selected using tournament selection on the set of archived individuals.
Out of the archived individuals, $P_{n}$ individuals are newly generated every generation.
By introducing asynchronous updates, we can achieve a high-performance using a massive parallel computer.

In our study, $P_{\rm ini} = 1000$, $P_{n} = 500$, and $P_{\rm archive} = 1000$ were used.
For each individual (i.e. input parameters of the simulator), we conducted five independent runs that had a different random number seed, and their results were averaged.
Simulated binary crossover~\cite{deb:sbx} and polynomial mutation~\cite{deb:moea} were used as genetic operators.
For the tournament selection parameters, a crossover rate of $1.0$, simulated binary crossover of $\eta_{b} = 15$, mutation rate of $0.01$, and polynomial mutation of $\eta_{p}=20$ were used.

\subsection{Evacuation Simulator}

To evaluate an evacuation plan, we used a multi-agent simulator CrowdWalk~\cite{yamashita:2013,yamashita:2014a}, which simulates the moves of pedestrians in a city.
The simulator adopts one-dimensional roads on which agents move; that is, the road network is represented by nodes and links.
This design is advantageous for making simulations sufficiently fast to manage a large number of agents.

\begin{figure}[htb]
  \centering
    \includegraphics[width=.9\textwidth]{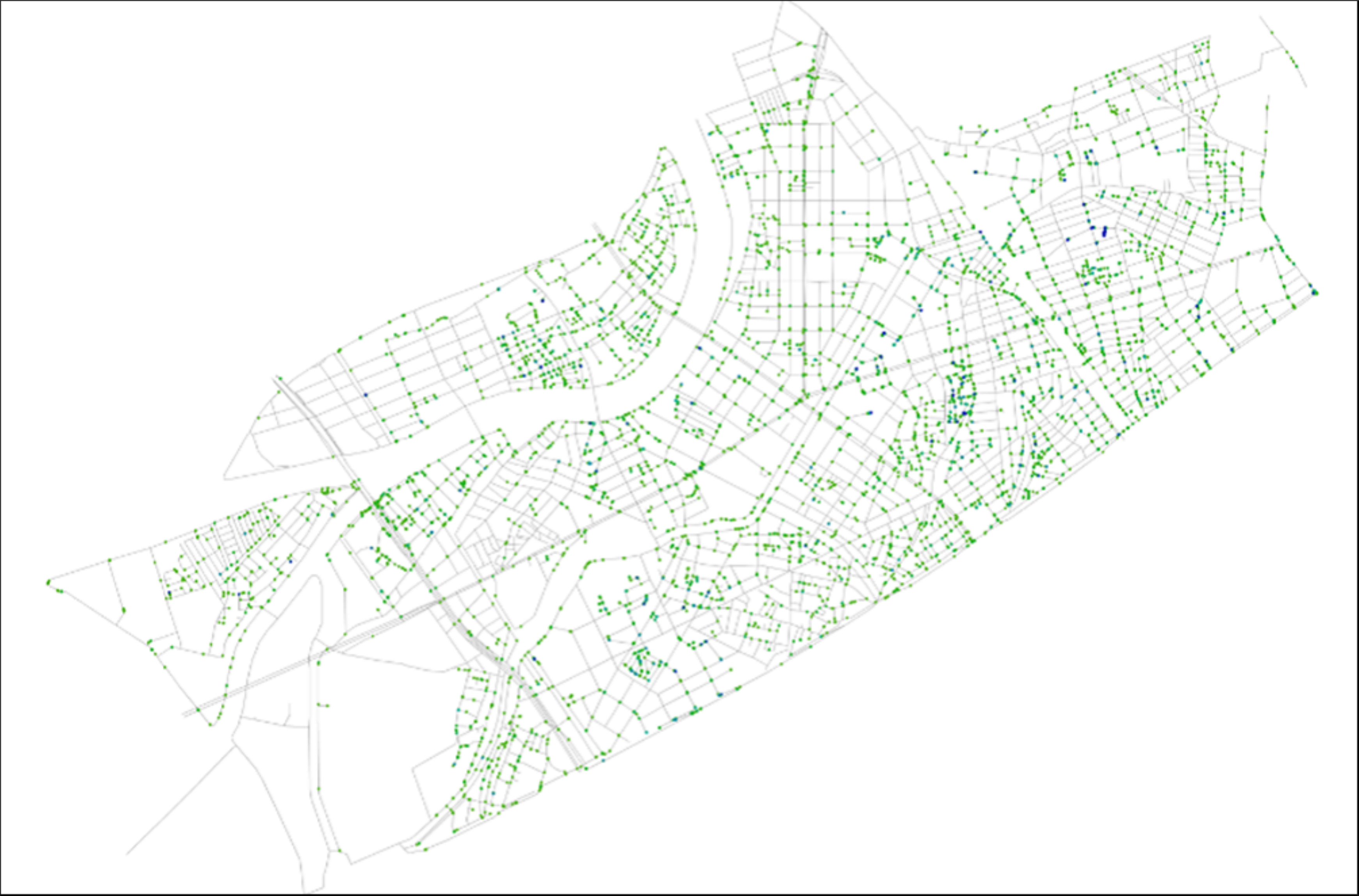}
    \caption{
    Snapshot of one of the evacuation simulations conducted in this study.
    The lines and green points indicate roads and agents, respectively.
    }
    \label{fig:crowdwalk}
\end{figure}

In this study, we simulated the evacuation of pedestrians in the Yodogawa district in Osaka, Japan.
The road network had 2,933 nodes and 8,924 links. In our setting, the number of evacuees and shelters were 49,726 and 86, respectively.
Figure~\ref{fig:crowdwalk} shows a snapshot of the simulation in this study.

In our study, the entire simulation area were divided into 533 sub-areas. Each sub-area had a given number of evacuees.
The evacuees in each sub-area were further divided into two groups in the ratio $r_i$ and $1-r_i$, where $i$ is an index of the sub-areas.
For each group, a shelter was assigned as an evacuation destination.
The ratios $r_i$ and destinations for each group are input parameters that characterize an evacuation plan.
Thus, we had 1,599 input parameters for this simulation as $\{r_i\}$ and two destinations were assigned to each sub-area.
We fixed other simulation parameters (e.g., the speed of the pedestrians) for simplicity.

We used the following three objective functions in this study:
\begin{description}
   \item[f1: time to complete the evacuation]\mbox{}\\
      Required time until all the agents arrive at their designated shelter. This is obtained from the simulation.
   \item[f2: complexity of the evacuation plan]\mbox{}\\
    We quantify the difficulty of the evacuation plan using the information entropy of the population distribution in each sub-area:\\
    $f_2 = \sum_i \left(r_i\log{(r_i)}+(1-r_i)\log{(1-r_i)}\right)$.
    If we do not split the residents in a sub-area into smaller groups, the evacuation plan becomes simpler. Thus, smaller entropy indicates a simpler evacuation plan.
    This quantity is calculated when an evacuation plan is given. 
   \item[f3: number of excess evacuees]\mbox{}\\
   This is a measure of the feasibility of a plan. Each shelter has a capacity, and the number of excess evacuees are measured.
   This quantity is calculated when an evacuation plan is given.
\end{description}
Solutions that minimize these objective functions were searched using NSGA-II.

\subsection{Results and Discussion}

We conducted an optimization on the K computer using 640 nodes and 5,120 CPU cores.
The population was updated for $40$ generations, and $105,000$ simulation runs were conducted in total.
Even though the elapsed time for each simulation run ranged significantly from 30 minutes to 50 minutes, depending on the simulation parameters, most of the simulation runs were conducted in parallel and their job balancing was good. The job filling rate achieved $93\%$ in our experiment.

\begin{figure}[!htb]
  \centering
    \includegraphics[width=.9\textwidth]{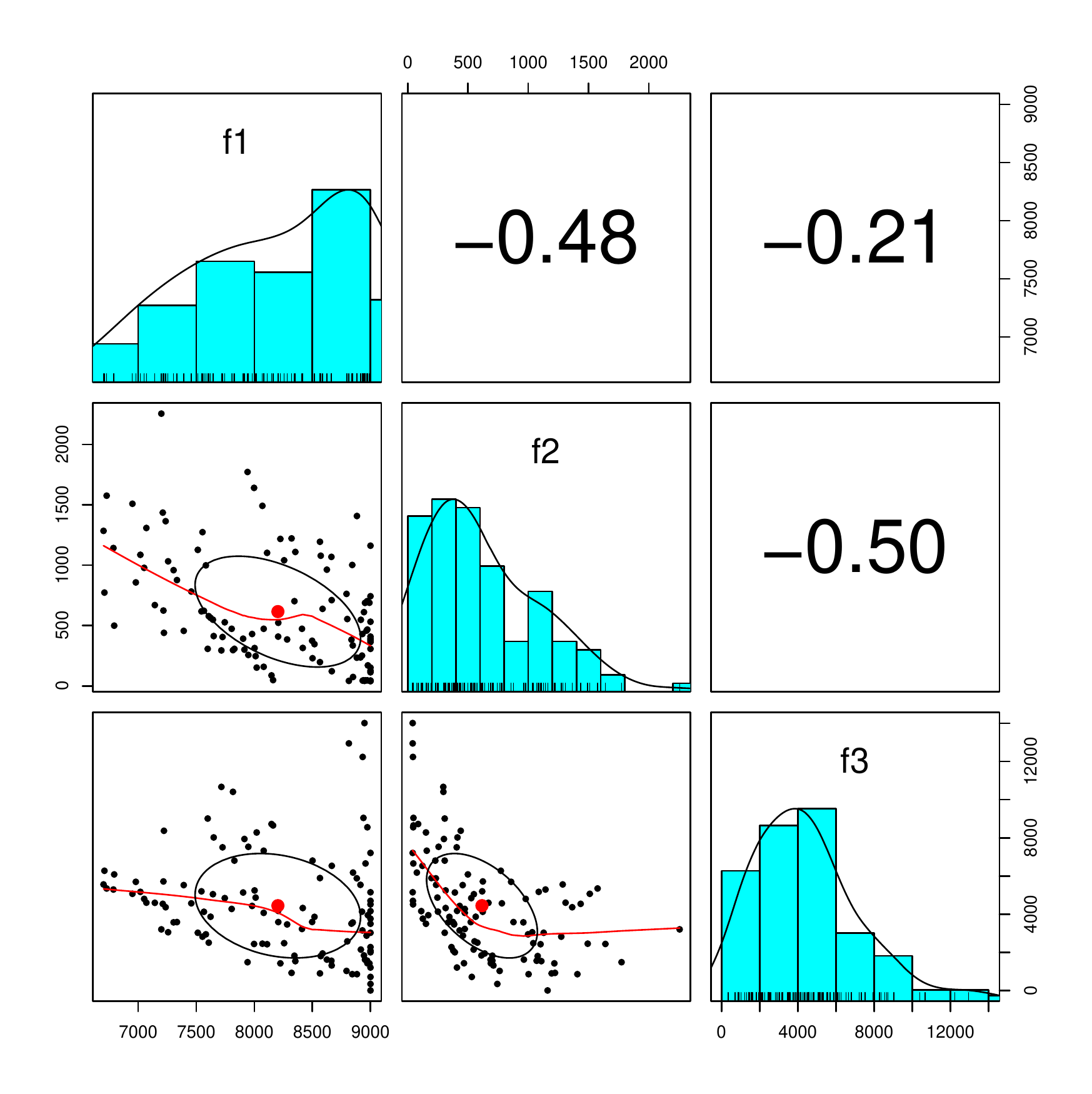}
    \caption{Solutions obtained after $40$ generations.
    In the left bottom panels, the solutions are shown in scatter plots, whereas their correlation coefficients are shown in the top right panels.
    In the diagonal panels, the histograms of the solution as a function of $f1$, $f2$, and $f3$ are shown.
    }
    \label{fig:correlation_front}
\end{figure}

Figure~\ref{fig:correlation_front} shows the solutions determined after $40$ generations.
In the left bottom panels of Fig.~\ref{fig:correlation_front}, scatter plots of the solutions on the Pareto front are shown.
Although they actually exist in three-dimensional space, they are mapped into two-dimensional spaces in these plots.
Clearly, there are negative correlations between a pair of the objective functions.
Their Pearson's correlation coefficients were calculated and are shown in the upper right panels.
In the diagonal panels, the histograms of the solutions are shown.
The correlation coefficients are negative, which indicates that there are trade-offs between these objective functions.
For instance, if we want to shorten the time for evacuation, a complex plan is needed.

\section{Conclusions and Future Work}

In this paper, we presented CARAVAN, a highly scalable framework for parameter-space exploration, which executes independent simulation runs in parallel on massive parallel computers.
Users can define a workflow using Python without any knowledge of MPI libraries, and the simulator can be implemented in an arbitrary language.
We evaluated the performance on the K computer and showed that it demonstrated good scaling for up to $16,384$ MPI processes.

As a case study, we applied the framework to an evacuation guidance problem.
When evaluating evacuation plans for a disaster scenario, there often exists a trade-off between the effectiveness and its implementation cost.
We demonstrated that CARAVAN is effective for solving this multi-objective optimization problem because it requires a large number of evaluations of plans using multi-agent simulations.

CARAVAN is an ongoing project, and its performance and usability will be improved in future releases.
In addition to these improvements, more studies on algorithms for search engines in massive parallel environments are strongly needed.
Most of the well-known algorithms for the design of experiments or optimizations assume the serial calculation of an objective function.
However, in our case, calculations of objective functions, that is, executions of a simulator, are conducted on highly parallel machines.
Effective algorithms for such a condition are expected to maximize the potential of MASS and HPCs.

\subsection*{Acknowledgement}
Y.M. acknowledges support from MEXT as ``Exploratory Challenges on Post-K computer (Studies of multi-level spatiotemporal simulation of socioeconomic phenomena)'' and the Japan Society for the Promotion of Science (JSPS) (JSPS KAKENHI; grant no. 18H03621).
This research used computational resources of the K computer provided by the RIKEN Center for Computational Science through the HPCI System Research project (Project ID:hp160264).
We thank Maxine Garcia, PhD, from Edanz Group (www.edanzediting.com/ac) for editing a draft of this manuscript.

%
%
%
%

\end{document}